# Investigation of the Chaotic Dynamics of an Electron Beam with a Virtual Cathode in an External Magnetic Field

**E. N. Egorov and A. E. Hramov**

*Saratov State University, Astrakhanskaya ul. 83, Saratov, 410026 Russia*

**Abstract**—The effect of the strength of the focusing magnetic field on chaotic dynamic processes occurring in an electron beam with a virtual cathode, as well as on the processes whereby the structures form in the beam and interact with each other, is studied by means of two-dimensional numerical simulations based on solving a self-consistent set of Vlasov–Maxwell equations. It is shown that, as the focusing magnetic field is decreased, the dynamics of an electron beam with a virtual cathode becomes more complicated due to the formation and interaction of spatiotemporal longitudinal and transverse structures in the interaction region of a vircator. The optimum efficiency of the interaction of an electron beam with the electromagnetic field of the vircator is achieved at a comparatively weak external magnetic field and is determined by the fundamentally two-dimensional nature of the motion of the beam electrons near the virtual cathode.

## 1. INTRODUCTION

Over the past three decades, considerable attention has been given to physical processes occurring in charged particle beams with an overcritical current in regimes of formation of a virtual cathode (VC) [1–6].

Numerous experimental and theoretical investigations of VC-based devices (which are usually called vircators [7]) give evidence of the complicated unsteady dynamics of radiation from beams with a VC in such systems [3, 8–12]. In particular, such investigations revealed some regular features of the chaotic generation of radiation in vircators [5, 6, 12–17], provided insight into the onset of spatiotemporal structures in an electron beam and on their interaction with a VC [16–21], and showed a close relationship between the complication of the spectral content of radiation from a vircator and the formation of electron structures in an overcritical-current electron beam [16, 18–20, 22].

However, the main regular features of the complicated nonlinear dynamic processes in VC-based devices have been examined primarily by using one-dimensional (1D) models of the electron beam dynamics. In a number of papers (see, e.g., [15, 17, 20, 22]), the electron beam was assumed to be fully magnetized, so transverse electron motions were ignored and the beam motion was treated as one-dimensional. The self-consistent electromagnetic field was modeled by numerically solving Poisson's equation (see, e.g., [15, 22, 23]) or Maxwell's equations [17, 20].

Of course, the assumption of 1D motion of an electron beam in a vircator is valid only for a sufficiently strong guiding magnetic field, but this is not always the case in a physical experiment. Moreover, a study was also made of experimental models of a vircator without an external focusing magnetic field [5, 9, 12, 24]. In addition, the vircator radiation spectra calculated in 1D models are in rather poor agreement with the data of physical experiments. In particular, it was shown experimentally that the bandwidth of chaotic generation in a vircator is $f/f_0 > 50\%$ [12], and, in some experiments, radiation was observed to be generated over more than an octave bandwidth [25, 26]. It can be supposed that this disagreement stems from the fact that, in 1D models, transverse electron motions in a beam focused by a weak (but finite) magnetic field are ignored. In [27, 28], attempts were made to theoretically study the processes in a beam with a VC by taking into account the finite strength of the magnetic field that focuses the beam electrons. However, the effect of the strength of the focusing magnetic field on the nonlinear unsteady processes that occur in an electron beam with an overcritical current has not yet been analyzed in detail.

In this paper, we present the results of numerical investigations of the nonlinear unsteady dynamic processes and structure formation in a relativistic electron beam (REB) with a VC. The investigations were carried out based on a 2.5-dimensional model in which the electron dynamics in the interaction space of the vircator is described by a particle-in-cell (PIC) method [4, 29, 30] and the self-consistent electromagnetic field is determined by numerically integrating Maxwell's equations in two dimensions by the finite difference method.

## 2. MATHEMATICAL MODEL

The vircator system under analysis is a portion of a cylindrical waveguide with the length $L = 0.16$ m and radius $R = 0.3L$. An axisymmetric annular monoenergetic REB with a zero transverse velocity and a uniform transverse current density distribution is injected into the working chamber through the cross section $z = 0$. The energy of the accelerated beam electrons is 560 keV (so the ratio of the electron velocity at the entrance to the chamber to the speed of light $c$ is equal to $v_0/c = 0.88$), the inner radius of the injected REB and the thickness of its wall being $r_b = 0.2L$ and $\Delta_b = 0.03L$, respectively. It is assumed that the electrons reflected from the VC are absorbed at the entrance grid; this corresponds to the reditron scheme of a vircator [31]. The external guiding magnetostatic field is produced by a cylindrical solenoid with a circular cross section and is assumed to be uniform within the working chamber of the vircator.

We will consider the dynamics of an electron beam in the actual complicated geometry of a vircator by using a time-dependent 2.5-dimensional model based on solving a self-consistent set of Vlasov–Maxwell equations [29, 32]. Note that such models are now being increasingly used in analyzing the physical process in various microwave electronic devices [33].

The general form of the model equations is as follows [29, 33, 34]:

$$\nabla \times \mathbf{E} = -\frac{1}{c}\frac{\partial \mathbf{H}}{\partial t}, \quad \nabla \times \mathbf{H} = \frac{1}{c}\frac{\partial \mathbf{E}}{\partial t} + \frac{4\pi}{c}\mathbf{j}, \qquad (1)$$

$$\nabla \cdot \mathbf{E} = 4\pi\rho, \quad \nabla \cdot \mathbf{H} = 0, \qquad (2)$$

$$\frac{\partial f}{\partial t} + \mathbf{v}\frac{\partial f}{\partial \mathbf{r}} + e\left(\mathbf{E} + \frac{1}{c}[\mathbf{v}, \mathbf{H}]\right)\frac{\partial f}{\partial \mathbf{p}} = 0, \qquad (3)$$

$$\rho = e\int f d\mathbf{p}, \quad \mathbf{j} = e\int f \mathbf{v} d\mathbf{p} \qquad (4)$$

where $\mathbf{E}$ and $\mathbf{H}$ are the electric and magnetic field strengths, $f = f(\mathbf{r}, \mathbf{p}, t)$ is the electron distribution function, $\rho$ and $\mathbf{j}$ are the charge and current densities, and $\mathbf{v}$ and $\mathbf{p}$ are the velocity and momentum of the electrons. Equations (1)–(4) are supplemented with the corresponding initial and boundary conditions.

We will be interest in systems with an axisymmetric interaction space. Such systems can be conveniently described in cylindrical coordinates such that $\partial/\partial\theta = 0$. Note that, with this simplification, axially asymmetric modes that can be excited in the interaction of a beam with an electromagnetic field are excluded from consideration [35].

In accordance with what was said above and under the assumption that the external alternating field components $E_\theta$, $H_r$, and $H_z$ are absent and the component $p_\theta$ of the initial electron beam momentum is zero, Maxwell's equations (1) reduce to

$$\begin{aligned} \frac{\partial H_\theta}{\partial t} &= -c\left(\frac{\partial E_r}{\partial z} - \frac{\partial E_z}{\partial r}\right), \\ \frac{\partial E_r}{\partial t} &= -c\frac{\partial H_\theta}{\partial z} - 4\pi j_r(r, z, t), \\ \frac{\partial E_z}{\partial t} &= \frac{c}{r}\frac{\partial (rH_\theta)}{\partial r} - 4\pi j_z(r, z, t). \end{aligned} \qquad (5)$$

Equations (5) with the corresponding initial and boundary conditions are integrated in a standard way on mutually staggered spatiotemporal numerical meshes with constant step sizes in time, $\Delta t$, and in the longitudinal and radial directions, $\Delta z$ and $\Delta r$. On each of the meshes, we will define one of the field components (see [29, 33, 34, 36] for details).

Vlasov equation (3) is solved by the PIC method. The electron beam is modeled by a finite number of macroparticles in the form of charged rings of different radii. For each of the macroparticles, the relativistic equations of motion

$$\frac{d\mathbf{p}}{dt} = \mathbf{E} + [\mathbf{p}, \mathbf{B}]/\gamma, \quad \frac{d\mathbf{r}}{dt} = \mathbf{p}/\gamma \qquad (6)$$

are solved by the Boris algorithm [37] adapted to cylindrical coordinates. The algorithm calculates the three components of the velocity of the charged macroparticle, namely, the longitudinal ($v_z$), radial ($v_r$), and azimuthal ($v_\theta$) components.

The space charge density $\rho(r, z, t)$ and current density $\mathbf{j}(r, z, t)$ of the electron beam are reconstructed by using a standard bilinear weighting procedure for macroparticles on a mesh in two dimensions in cylindrical coordinates [29].

## 3. INVESTIGATION OF THE COMPLICATED UNSTEADY DYNAMICS OF AN REB WITH A VIRTUAL CATHODE IN MAGNETIC FIELDS OF DIFFERENT STRENGTHS

Here, we consider the nonlinear dynamics of an electron beam with a VC at different strengths $B$ of the guiding magnetic field.

Figure 1 shows the oscillation spectra of the longitudinal electric field in the VC region of an electron beam with the current $I = 1.5I_0$ (where $I_0$ is the limiting vacuum current above which an unsteady VC forms in the beam) at different magnetic field strengths $B$.

In a strong magnetic field in which the electron beam dynamics is essentially one-dimensional, the oscillations excited in the system are close to single-frequency ones (see Fig. 1a, calculated for $B = 40$ kG). The oscillation spectrum has a pronounced, predominating peak at a frequency of $f \approx 2.5(\omega_p/2\pi)$. This result agrees with the predictions of the steady-state theory

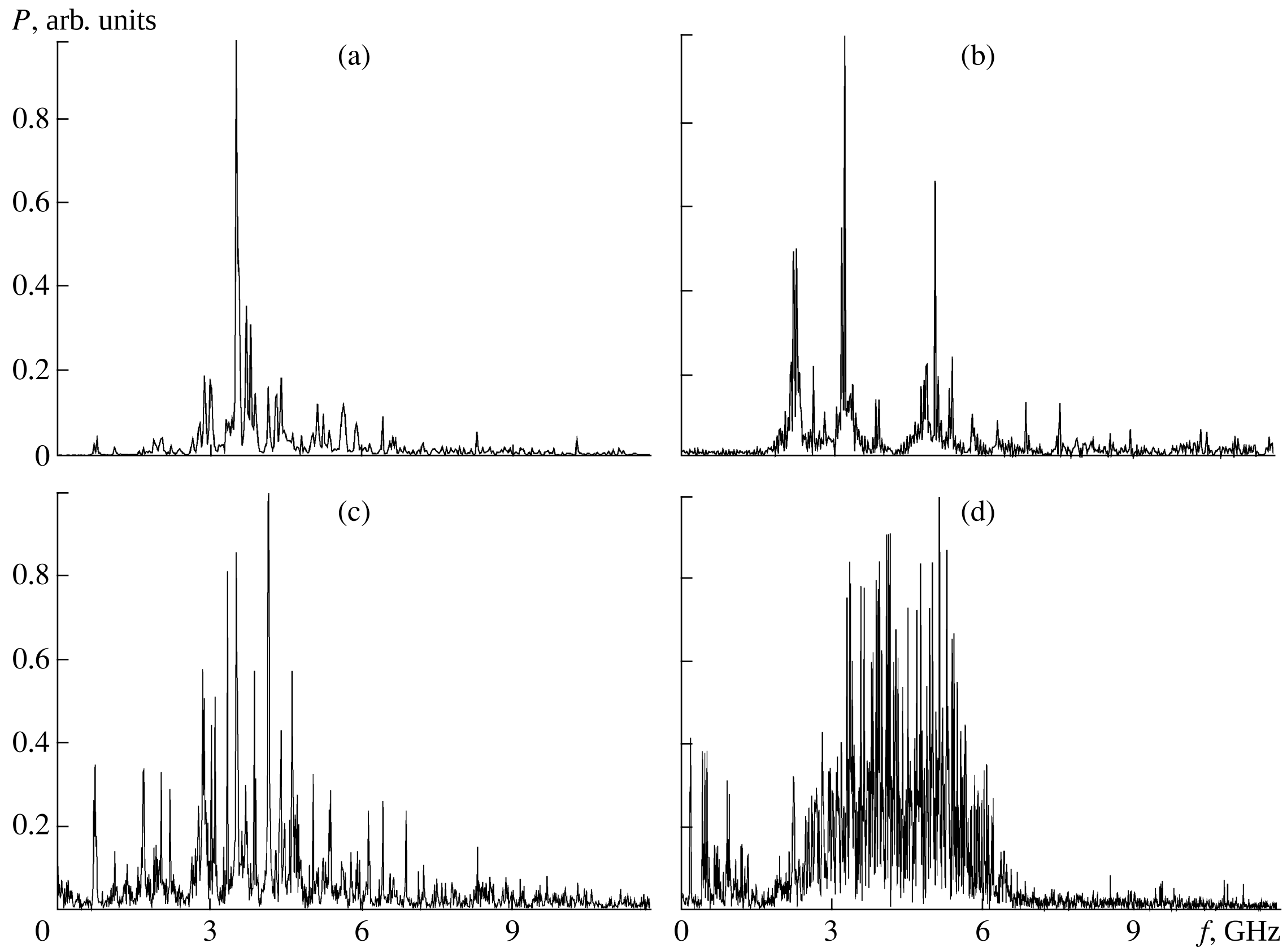

**Fig. 1.** Normalized power spectra of the electric field oscillations in the VC region for different strengths of the guiding magnetic field: $B$ = (a) 40, (b) 20, (c) 6, and (d) 0 kG.

[38], according to which the frequency $\omega_{VC}$ of VC oscillations is estimated to be $1.9 \lesssim \omega_{VC}/\omega_p \lesssim 2.5$.

At weaker guiding magnetic fields, the generation process is more complicated: the oscillation spectrum has additional frequency harmonics and a higher noisy pedestal at a level from –30 to –20 dB (see Fig. 1b, calculated for $B$ = 20 kG). At weak fields $B$, the spectrum has not only high-frequency (HF) harmonics at 3–6 GHz but also low-frequency (LF) components at frequencies $f$ < 500 MHz (see Fig. 1c, calculated for $B$ = 6 kG). In the absence of a magnetic field, the oscillation power spectrum in the frequency range 3.5–6 GHz is continuous, with an even higher noisy pedestal at a level of –10 dB (see Fig. 1d). The degree to which the spectrum is irregular, $N = P_{max}/P_{min}$ (where $P_{max}$ and $P_{min}$ are the maximum and minimum powers of the spectral harmonics in the working frequency range), does not exceed 3–10 dB.

In studying two-dimensional effects in the dynamics of an electron beam with a VC, it is important to analyze the influence of the current distribution in the injected beam on the current $j_{tr}$ that is transmitted through the VC and reaches the collector end of the interaction space, on the current $j_r$ that is reflected from the VC back toward the injection plane, and on the current $j_w$ to the side wall that bounds the drift space in the transverse direction. Figure 2 shows the dependence of the time-averaged currents $j_{tr}$, $j_r$, and $j_w$ on the strength $B$ of the focusing magnetic field at a low and at a high beam current. The currents $j_{tr}$, $j_r$, and $j_w$ are normalized to the current of the injected beam. In Fig. 2, we also show the gyrofrequency $\omega_c$ of the beam electrons in the focusing magnetic field. From Fig. 2 we can see that, at strong focusing magnetic fields ($B$ > 10 kG), the currents $j_{tr}$ and $j_r$ depend weakly on the field strength. As $B$ is increased, the current $j_{tr}$ grows insignificantly and the current $j_r$ falls off accordingly. In this range of $B$ values, the two-dimensional electron dynamics in a finite magnetic field does not change the current distribution in the system. As in the case of an infinitely strong magnetic field (when the electron dynamic is inherently one-dimensional), the injected electron beam in the VC region is dynamically distributed into two streams—one that is reflected from the VC back toward the injection plane and one that is transmitted through the VC and reaches the collector.

Note also that, at strong focusing magnetic fields, the transmitted current $j_{tr}$ is close in magnitude to the limiting vacuum current in the interaction space. This result agrees well with the known experimental and the-oretical data (see, e.g., [39, 40]), according to which the current transmitted through the VC should be very close to the limiting vacuum current. The dependence of the transmitted current on the injected beam current at the magnetic field $B = 60$ kG is presented in Fig. 2c. We can see that, at low injected currents, the transmitted current is lower than the limiting one. As the beam current is increased, the transmitted current rapidly approaches the limiting vacuum value. The dependence of the transmitted current on the injected current requires further theoretical analysis. From Fig. 2c we can see that, at weak focusing magnetic fields, the transmitted current is lower than the limiting vacuum current because some of the beam electrons escape to the side wall that bounds the drift space in the transverse direction. As the focusing magnetic field is increased, the fraction of beam electrons that escape to the side wall decreases and, accordingly, the transmitted current approaches its limiting vacuum value.

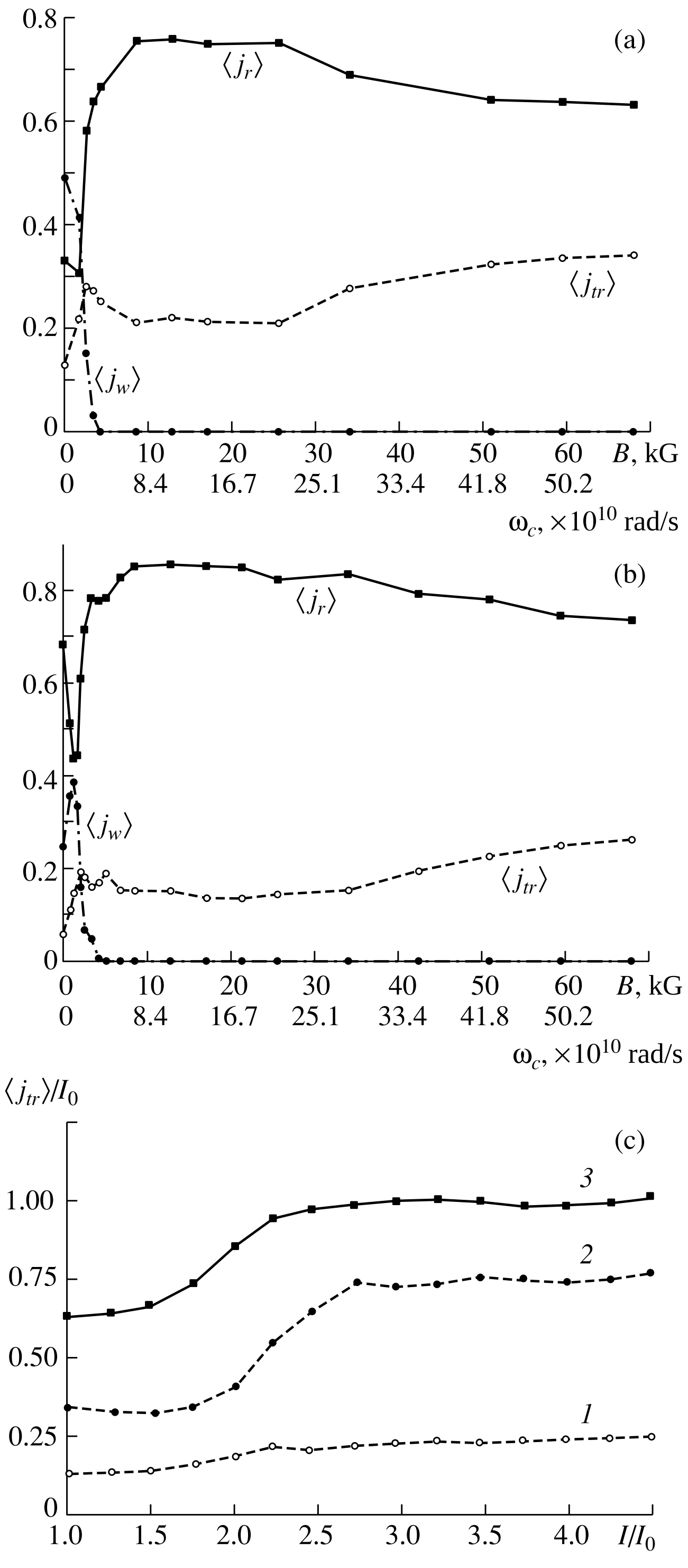

**Fig. 2.** Time-averaged currents of the electrons reflected from the VC toward the injection plane, the electrons that are transmitted through the VC and reach the collector, and the electrons that escape to the side wall as functions of the magnetic field $B$ (and the electron gyrofrequency $\omega_c$) for $I$ = (a) $1.5I_0$ and (b) $4I_0$. (c) Transmitted current vs. injected current for different magnetic field strengths: $B$ = (*1*) 60, (*2*) 4, and (*3*) 0 kG.

At weak focusing magnetic fields, the beam dynamics is radically different. The beam current begins to be absorbed by the side wall of the working chamber of the vircator at magnetic fields as weak as $B < 5$–10 kG, and the absorbed fraction of the beam current grows rapidly as the magnetic field is decreased. The current $j_w$ is maximum at $B \sim 0$–2 kG. At such magnetic fields, the current escaping from the system through the injection plane is very low. An analysis of the electron trajectories shows that the beam electrons absorbed at the side wall are mostly those that have been reflected from the VC. The magnetic field $B_w$ at which the beam electrons begin to be absorbed by the side wall can be regarded as a critical field strength at which two-dimensional effects begin to considerably affect the space charge dynamics. It is in the magnetic field range $B < B_w$ that the multifrequency chaotic signal from the vircator is observed to be generated over a substantially broader frequency band.

Hence, as the focusing magnetic field is decreased while the beam current is held fixed, the frequency spectrum generated in a system with a VC becomes progressively more complicated: it changes from a single-frequency spectrum, when the electron dynamics is essentially one-dimensional, to the broadband spectrum of multifrequency chaotic oscillations, when the beam dynamics is inherently two-dimensional and the electrons are observed to be absorbed at the side wall.

The general conclusion of this section is as follows. As the guiding magnetic field is decreased (and, accordingly, as the two-dimensional electron dynamics exerts a progressively greater effect on the oscillations generated in a system with a VC), the output radiation spectrum becomes more complicated. In this context, an important practical task is to examine how the efficiency of a VC-based system depends on the strength of the guiding magnetic field.

## 4. EFFICIENCY OF A VC-BASED OSCILLATOR AS A FUNCTION OF THE MAGNETIC FIELD STRENGTH

The efficiency $\eta$ of the interaction of an electron beam with an electromagnetic field in a vircator is defined as

$$\eta = \frac{W_{e\,\mathrm{in}} - W_{e\,\mathrm{out}}}{W_{e\,\mathrm{in}}}, \qquad (7)$$

where $W_{e\mathrm{in}}$ is the electron beam energy injected into the interaction space and $W_{e\mathrm{out}}$ is the energy of the electrons that escape from the interaction space. It is easy to show that the interaction efficiency (7) can be expressed by the relationship

$$\eta = \frac{\gamma_0 - \frac{1}{M}\sum_{k=1}^{M}\gamma_k}{\gamma_0 - 1}, \qquad (8)$$

where $\gamma_0$ is the relativistic factor of the beam electrons at the entrance to the interaction space, $\gamma_k$ is the relativistic factor of the $k$th charged particle that leaves the interaction space, and $M$ is the total number of charged particles that have escaped from the interaction space.

Figure 3 illustrates how interaction efficiency (8) depends on the strength $B$ of the focusing magnetic field for different ratios of the beam current to the limiting vacuum current $I_0$ (above which an unsteady VC forms in the system). We can see that, for each value of the beam current, there exists an optimum magnetic field strength $B_{\mathrm{opt}}$ at which the interaction efficiency is highest. The optimum magnetic field is weak, $B_{\mathrm{opt}} \lesssim$ 10 kG, and increases with the beam current. For instance, for $I \approx I_0$, we have $B_{\mathrm{opt}} \approx 0$, whereas, for $I = 4I_0$, the optimum magnetic field is as strong as $B_{\mathrm{opt}} \approx 9$ kG. For the beam parameters considered in this section, the maximum vircator efficiency is about $\eta \sim 0.06$.

When the guiding magnetic field is changed with respect to its optimum strength, the efficiency of the interaction of a beam with an electromagnetic field decreases abruptly. In the range of strong magnetic fields, $B > 20$ kG, the interaction efficiency varies only slightly and does not exceed values of $\eta \sim 0.01$–$0.02$, depending on the relative beam current $I/I_0$.

The above theoretical formulas for the dependence of the generation efficiency of a vircator on the strength of the guiding magnetic field agree well with the experimental data known from the literature. Davis et al. [31] revealed experimentally that an optimum magnetic field (of comparatively low strength) at which the generation power in a VC-based system is highest does indeed exist, but no physical explanation was offered.

Note that formula (7) for the efficiency of the interaction of a beam with a field in a vircator accounts for only the energy of the electrons that are injected into the interaction space and escape from it. In a number of papers (see, e.g., [12, 22]), it was shown that, in some of the operating modes of a vircator, the metastable particles that remain in the interaction space for a long time (much longer than the oscillation period in the system) begin to play an important role in the dynamics of the VC. In this case, formula (7) should be supplemented with a term that takes into account the energy of the charged particles that remain in the working chamber. With formula (7) so refined, we calculated the interaction efficiency for the beam current $I = 1.5I_0$. For an optimum magnetic field $B$, the refined efficiency was calculated to be $\bar{\eta} = 0.068$, whereas the efficiency calculated for the same magnetic field but without allowance for the metastable particles was found to be equal to $\eta = 0.063$. In other words, when the metastable particles remaining in the working chamber are taken into account, the interaction efficiency increases by a relative amount of about $\delta = |\bar{\eta} - \eta|/\eta \approx 0.08$. As the field $B$ is increased, the value of $\delta$ decreases; for strong magnetic fields ($B > 20$ kG), the calculated value of the efficiency is independent of whether the metastable particles are taken into account or not.

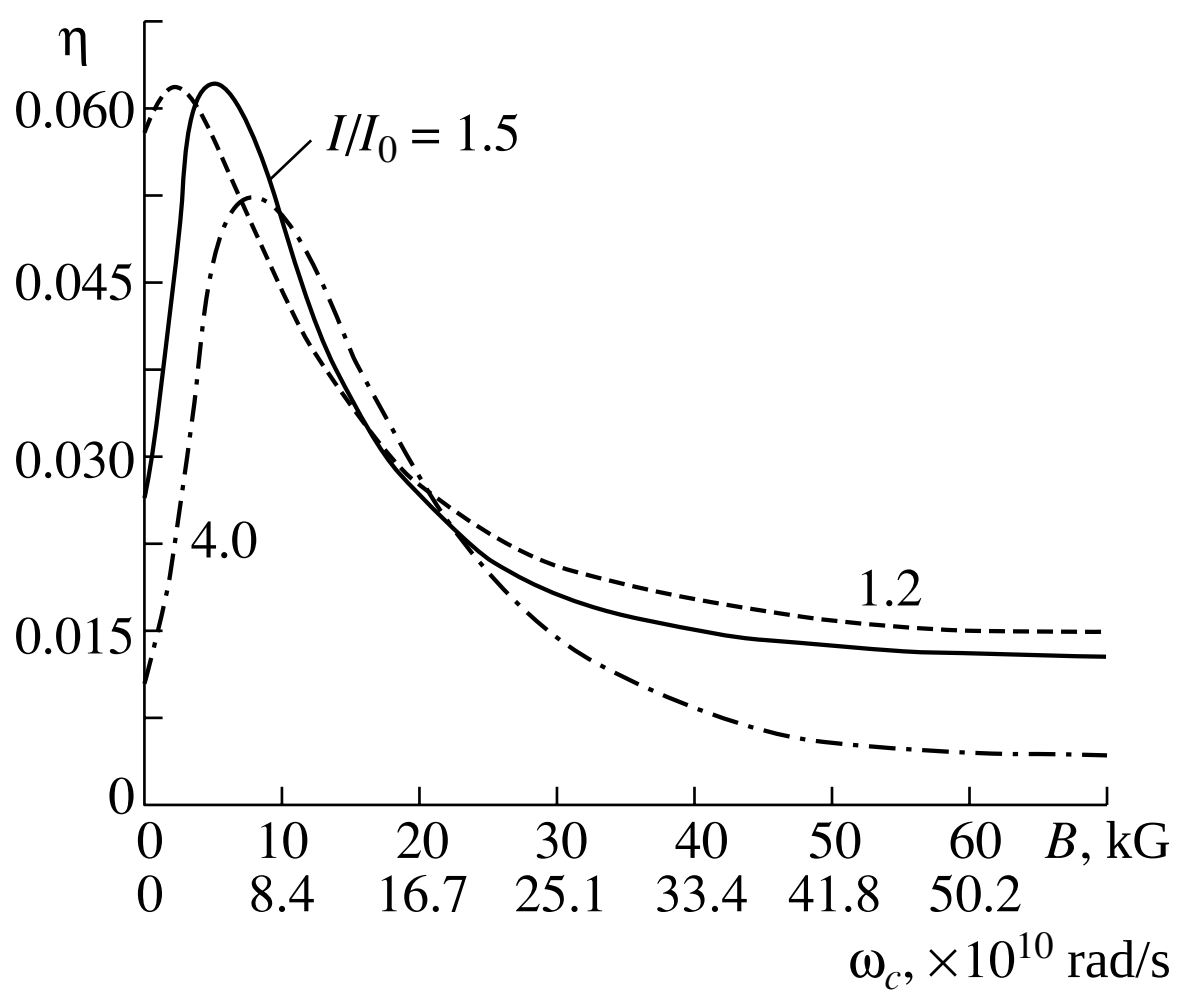

**Fig. 3.** Interaction efficiency vs. focusing magnetic field (electron gyrofrequency $\omega_c$) for different ratios of the beam current $I$ to the limiting vacuum current $I_0$.

Such a dependence of the parameter $\delta$ on the strength of the focusing magnetic field shows that the number of metastable particles increases substantially as the field is reduced. On the other hand, the effects associated with metastable particles do not appreciably change the value of the interaction efficiency; they merely introduce corrections to the results of calculations based on formula (7) or (8). How the number of metastable particles in the working chamber depends on the magnetic field strength will be discussed in more detail in the next section.

Let us now consider the reasons for such a dependence of the interaction efficiency on the strength of the focusing magnetic field. To do this, we analyze the space charge density distribution $\rho(r, z)$ of the electron beam and the electric field distribution $E(r, z)$ in the VC region. A fundamentally important point is that we must distinguish between the electrons that move toward the VC (their velocity is positive, $v_z > 0$) and the electrons the are reflected from the VC back toward the injection plane (their velocity is negative, $v_z < 0$). The first group of beam electrons (namely, the electrons that are injected into the interaction space) is decelerated by the electric field $E_z$ of the potential barrier of the VC; the velocity of these electrons decreases and they transfer their energy to the HF field. In contrast, the second group of electrons (namely, the reflected electrons) escapes from the VC region and is accelerated by the same field $E_z$, so the electrons take energy from the HF field. In this case, the field $E_z$, which decelerates the injected electrons and simultaneously accelerates the reflected electrons (whose velocity is negative, $v_z < 0$) is a function of the radial position $r$ and has a maximum at a radius $r_b$ equal to the radius of the injected beam, $\max E_z(r) = E_z(r = r_b)$. For $r > r_b$ and $r < r_b$, the electric field strength is a rapidly decreasing function. This is demonstrated in Fig. 4, which shows a qualitative characteristic profile of the electric field $E_z$ in the radial direction $r$ within the VC region in the interaction space. The shape of the profile $E_z(r)$ is essentially the same for a strong and a weak guiding magnetic field.

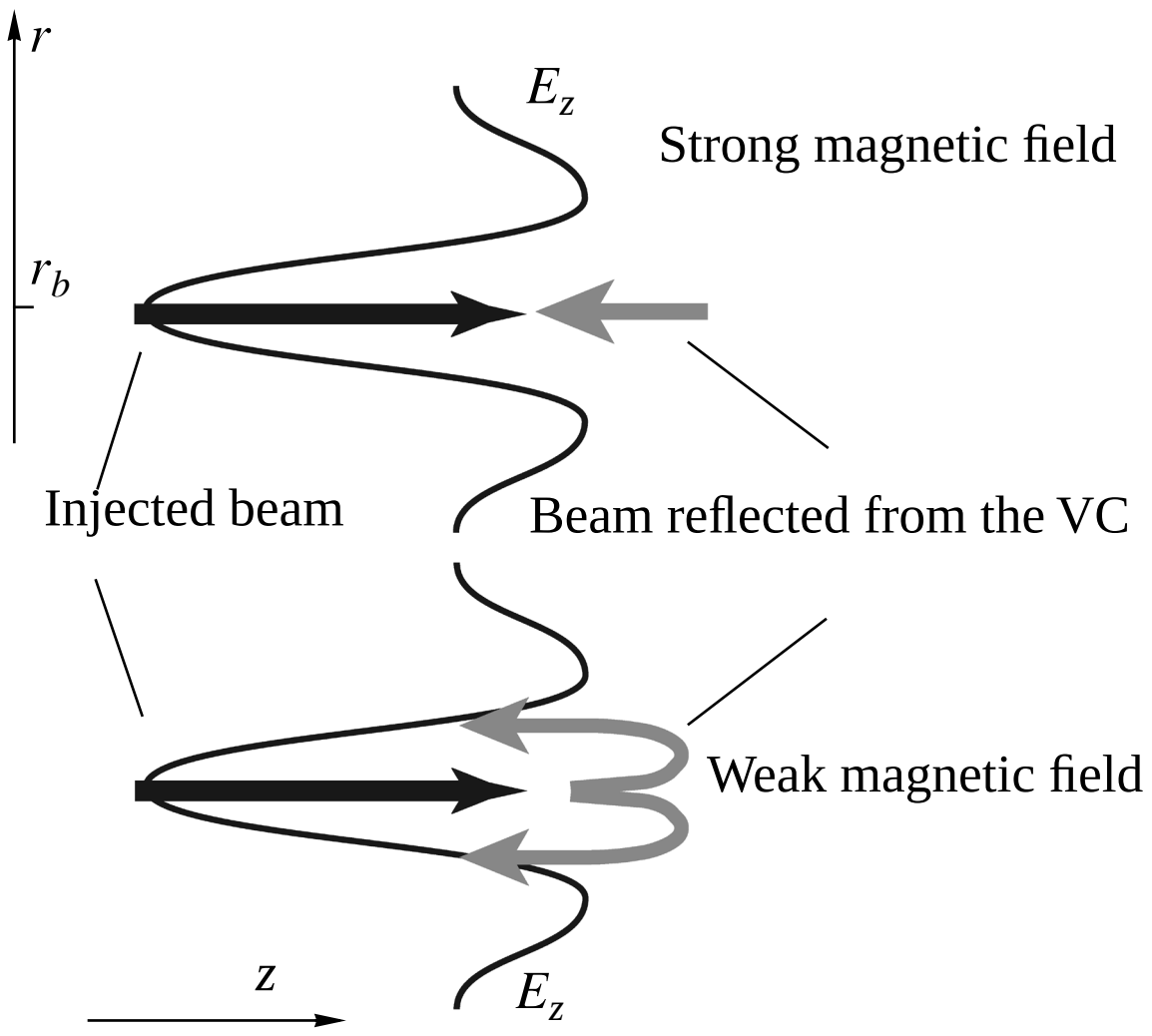

**Fig. 4.** Qualitative pattern of the beam dynamics in the VC region in a strong and in a weak magnetic field.

Figure 4 also qualitatively illustrates the characteristic features of the electron dynamics in the VC region at a strong (top) a weak (bottom) focusing magnetic field. In the coordinate plane $(r, z)$, the black heavy arrows show typical electron trajectories of the injected electron beam, which is decelerated by the electric field $E_z$. The gray arrows show typical trajectories of the electrons that are reflected from the VC and, accordingly, are accelerated by the field $E_z$. In a strong guiding magnetic field, the electron beam reflected from the VC has the same radius as the injected electron beam. Consequently, the reflected electrons are affected by the electric field that has the same amplitude as that influencing the injected electrons. As such, the energy transferred from the injected beam electrons to the HF field during their deceleration is close to the energy acquired by the reflected electrons in their acceleration. The result is that the interaction efficiency (and accordingly, the radiation power) turns out to be very low, $\eta \sim 0.01$–$0.02$.[1]

In a weak magnetic field, the situation is quite different. Since the reflected beam is focused only slightly, it can return to the injection plane along a trajectory different from that of the injected beam. The lower sketch in Fig. 4 qualitatively illustrates the characteristic trajectories of the reflected electrons in a weak magnetic field $B$. It is clearly seen that, since the reflected electrons move along trajectories other than those of the injected electrons, they experience an accelerating field that is weaker than the decelerating electric field. Therefore, the energy the accelerated reflected electrons take from the HF field is lower than that transferred from the injected electrons to the field. As a result, as the focusing magnetic field is decreased, the interaction efficiency increases and, at an optimum magnetic field, becomes as high as $\eta \sim 0.06$.

Note that, at a fixed energy of the injected electron beam, the higher the injection current and, accordingly, the greater the space charge, the stronger the defocusing effect of the latter. Consequently, as the injection current is increased, the optimum magnetic field should also be raised in order for the conditions for the interaction of a beam with the field to be as efficient as possible (see Fig. 4).

Let us now consider the quantitative parameters of the beam dynamics in the VC region. Figure 5 presents radial profiles of the longitudinal electric field, the electron beam space charge density, the space charge density of the electrons moving toward the VC region, and the space charge density of the electrons reflected from the VC. The profiles were obtained from simulations carried out for both weak and strong guiding magnetic fields and refer to different cross sections of the drift space in the region between the injection plane and the VC.

An analysis of the profiles in different cross sections $z$ of the interaction space shows that, at a strong focusing magnetic field, the space charge density of the beam electrons reflected from the VC is maximum in the same region as that of the beam injected into the inter-action space. Consequently, the beam electrons that have been reflected from the VC and escape from the interaction space through the injection plane experience an accelerating electric field of the same strength as that decelerating the injected electrons. In contrast, from Fig. 5b we see that, at a weak focusing magnetic field, the space charge density of the beam electrons reflected from the VC is maximum at a weaker longitudinal electric field. As a result, the energy the reflected electrons take from the electromagnetic field is lower than that transferred from the injected beam to the field and, accordingly, the interaction efficiency increases substantially, from 0.01–0.02 to 0.06.

[1] Note that, in this qualitative analysis, the electrons that have passed through the VC are ignored. Such electrons are also accelerated and, hence, take energy from the HF field. But the total number of them is much less than the number of electrons reflected from the VC (see Fig. 2) and they can be ignored in a qualitative analysis.

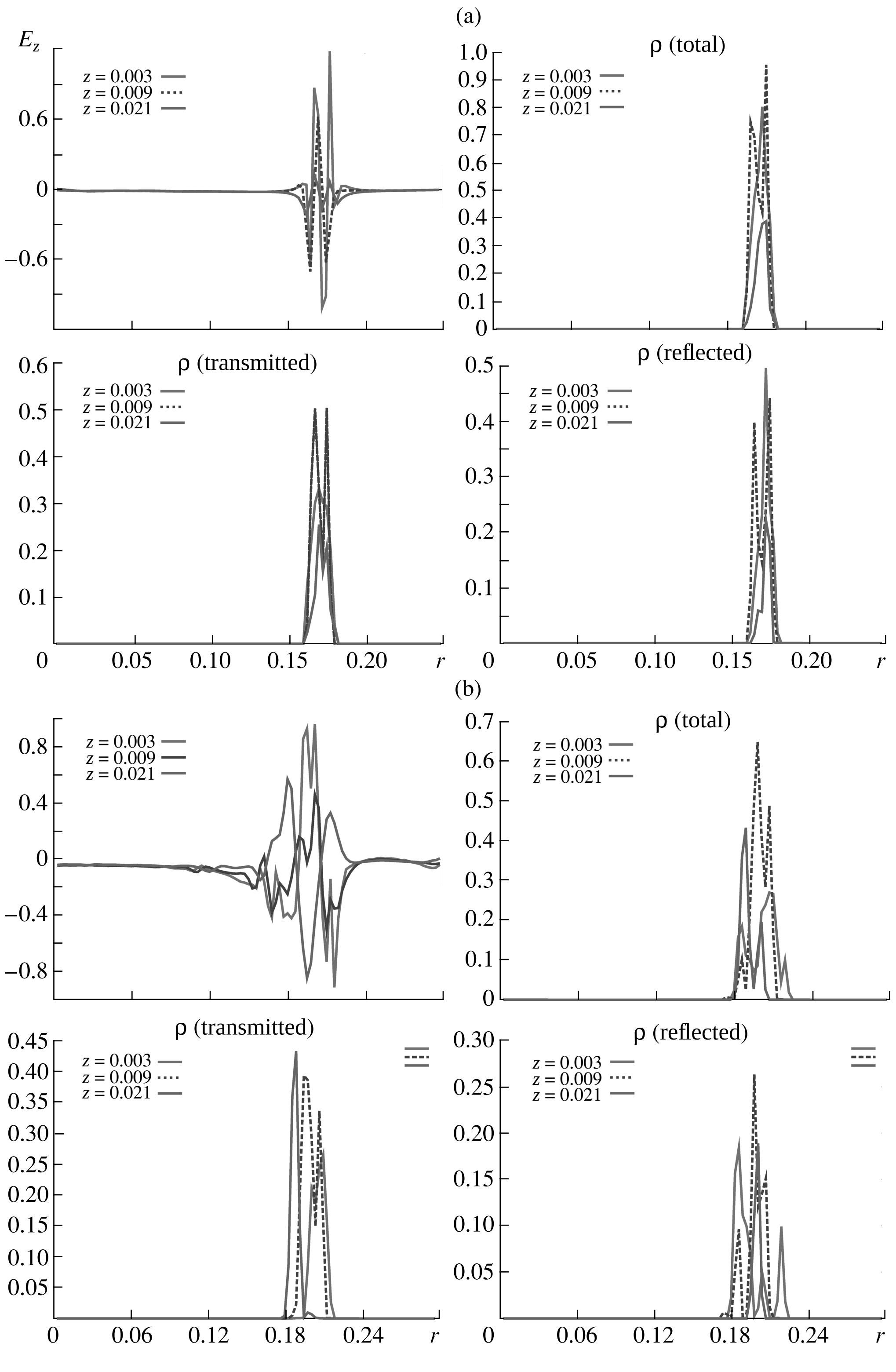

**Fig. 5.** Radial profiles of the longitudinal electric field, the total space charge of the electron beam, the space charge of the beam electrons moving toward the VC, and the space charge of the beam electrons reflected from the VC (a) in a strong magnetic field ($B$ = 40 kG) and (b) without a magnetic field ($B$ = 0).

## 5. INVESTIGATION OF THE STRUCTURE FORMATION PROCESSES IN AN REB

Here, we consider the physical processes whereby the chaotic dynamics in an electron beam with a VC becomes more complicated as the magnetic field is decreased. It is known that one of the physical mechanisms responsible for the complex beam dynamics in a system with an overcritical current is associated with the formation and interaction of spatiotemporal structures in a beam with a VC. This is why it is very important to study the dynamics of the space charge of an electron beam with an overcritical current in the interaction space in order to reveal the internal structures that govern the operating modes of a vircator system. Investigation of the structure formation processes should begin with the singling out of coherent structures that indicate the presence of internal motions responsible for chaotization of the VC dynamics in the system. In [22], it was proposed to single out the coherent structures by a method based on the wavelet bicoherence analysis of the spatiotemporal data. In this way, coherent structures can be conveniently singled out by using the Karunen–Loev (KL) orthogonal decomposition of the spatiotemporal data [20]. The problem of singling out the KL modes reduces to that of solving the integral equation

$$\int K(z, z')\Psi(z')dz' = \Lambda\Psi(z), \quad (9)$$

the kernel of which has the form

$$K(z, z') = \xi(z, t)\langle\xi(z', t)_t\rangle, \quad (10)$$

where the symbol $\langle\ldots\rangle_t$ stands for the time averaging. It is possible to choose a zero-mean set of spatiotemporal distributions of any physical quantity as the function $\xi(z, t)$. In our analysis, the function $\xi(z, t)$ can be conveniently taken to be the beam current density $j_z(z, t)$. The eigenvalue $\Lambda_n$ corresponding to the $n$th KL mode $\Psi_n$ is proportional to the energy contained in this mode. The measure of this energy can be characterized by the quantity

$$W_n = \Lambda_n/\sum_i \Lambda_i(\times 100\%). \quad (11)$$

In the situation in question, however, we deal with a two-dimensional electron motion; consequently, it is convenient to single out "longitudinal" and "transverse" structures in the beam separately. In the first case (the singling out of transverse structures, which characterize the radial beam space charge dynamics), the kernel of Eq. (9) is equal to

$$K(r, r') = 1/T 1/L\int_0^T\int_0^L u(z, r, t)u(z, r', t)dzdt. \quad (12)$$

In the case when longitudinal structures are singled out (i.e., those that characterize the dynamics of the system in the beam propagation direction), the kernel of the equation has the form

$$K(z, z') = 1/T 1/R\int_0^T\int_0^R u(z', r, t)u(z', r, t)drdt. \quad (13)$$

Here, $u(z, r, t)$ is a zero-mean set of spatiotemporal data (we used a set of values of the beam space charge density $\rho(z, r, t)$ minus its mean value) and $T$ is the time interval over which the time averaging is performed. As above, the energy of the transverse and longitudinal coherent structures is determined by analyzing the eigenvalues corresponding to the KL modes in Eq. (9).

Figure 6 displays how the energy $W_n$ (in percentage terms) of five highest-energy transverse and longitudinal coherent structures depends on the focusing magnetic field strength (the gyrofrequency $\omega_c$) for the beam current $I = 1.5I_0$. Note that, in all operating modes (from modes with regular oscillations to those with a complex chaotic beam dynamics), the number $n$ of coherent structures whose energy exceeds 90% of the total oscillatory energy is $n$ = 2–5.

At strong magnetic fields $B$, the energy of the oscillatory motion in the system is mostly concentrated in two transverse and one longitudinal KL structures. As the magnetic field is decreased, new structures arise in the system, with the result that the energy of the highest energy structures decreases and the amount of energy in the lowest modes with the numbers $n$ = 2–5 grows. At a weak magnetic field ($B \sim 0$), the transverse (radial) dynamics of the system is dominated by the five lowest modes, having nearly the same relative energy. This circumstance provides evidence that the space charge dynamics is complicated just in the transverse direction. As for the longitudinal structures, only one of them, namely, the main structure with the highest energy $W_1$, predominates even at a weak magnetic field; it is the structure that should be associated with the VC dynamics. This result agrees well with the results obtained in [17] in analyzing the structure formation processes in an electron beam with a VC in a one-dimensional model.

In order to confirm the relationship between the complication of the electron beam dynamics and the dynamics of the structures, we calculated the distributions $\Phi(\tau)$ of charged particles over their residence times in the interaction space. The results of these calculations are illustrated in Fig. 7. From Fig. 7a we can see that, for a strong magnetic field, the distribution function has one pronounced maximum (indicated by the symbol $T_{VC}$), which corresponds to the oscillation period of the VC (at the fundamental frequency $f = 1/T$ in the power spectrum that was calculated for the same values of the beam current and of the magnetic field; see Fig. 1a). We can conclude that, in this case, the beam dynamics is dominated by only one structure—the VC, so a vircator system operating with a strong focusing magnetic field generates nearly single-frequency radiation. This conclusion is confirmed by Fig. 8a, which shows configurational patterns of an electron beam for the magnetic field $B$ = 40 kG. The points in the patterns correspond to charged macroparticles used in numerical simulations. From the configurational pattern in the ($z$, $v_z/c$) coordinates in Fig. 8a, we can see that the bulk of charged particles are reflected from the same point $x = x_{VC}$ in the interaction space, i.e., from the VC (shown by the vertical dashed line). The configurational pattern in the ($z$, $r$) coordinates in Fig. 8a shows that the thickness of the wall of an electron beam propagating along the interaction space does not change. Hence, in a vircator system operating with a strong guiding magnetic field, only one electron structure (a VC) forms, which governs the dynamics of the system in the longitudinal direction.

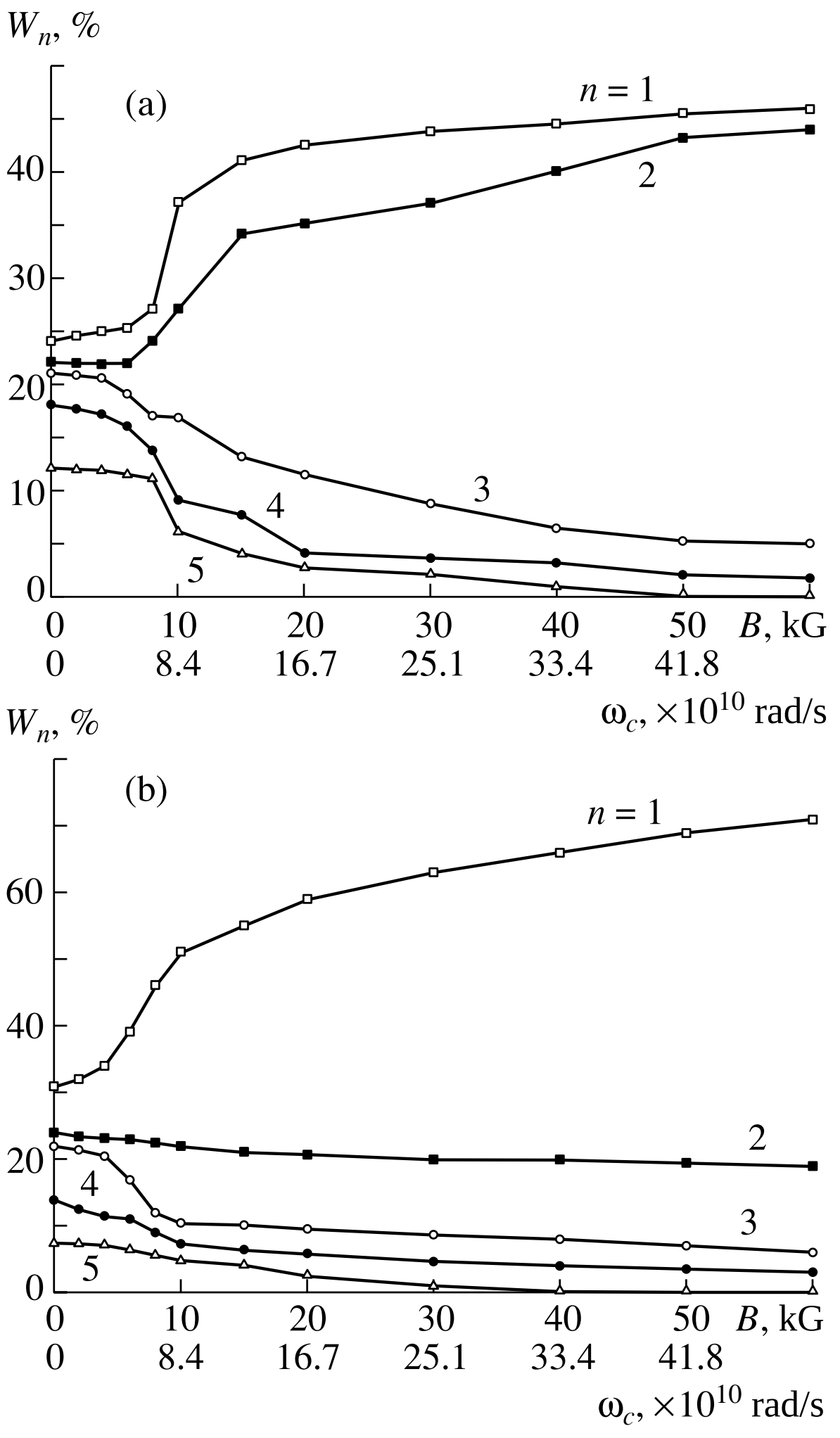

**Fig. 6.** Highest energy $W_n$ of the (a) transverse and (b) longitudinal coherent structures vs. focusing magnetic field (electron gyrofrequency $\omega_c$).

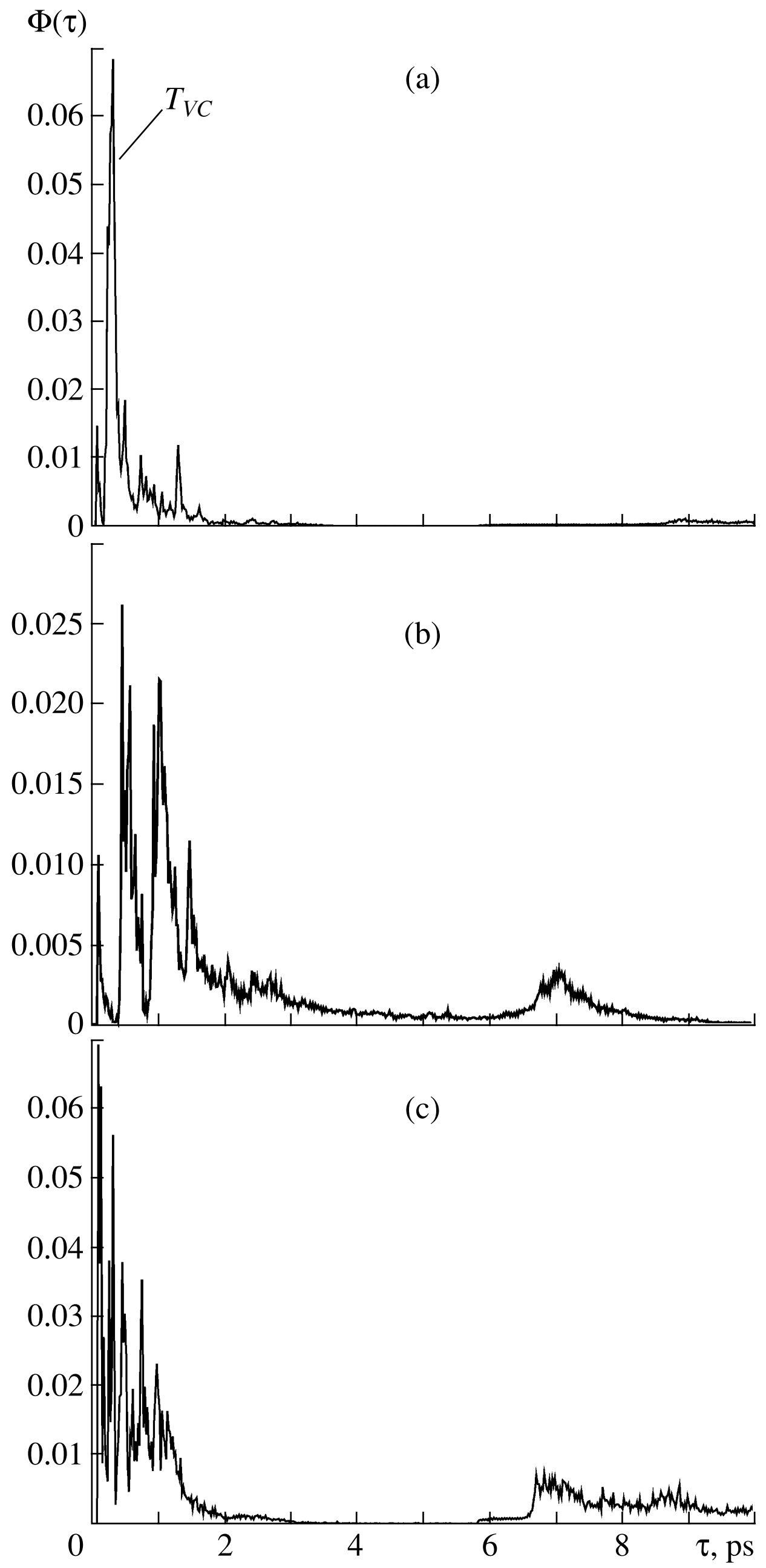

**Fig. 7.** Distributions of charged particles over their residence times in the interaction space for $B$ = (a) 40, (b) 8, and (c) 0 kG.

As the magnetic field is decreased (see Figs. 7b, 7c), the distribution function $\Phi(\tau)$ becomes more complicated in shape. Now, the function $\Phi(\tau)$ has several peaks, each corresponding to a different electron struc-ture in the beam. An analysis of the configurational patterns of an electron beam calculated for weak guiding magnetic fields (see Figs. 8b, 8c) also shows that the electron dynamics in the VC region becomes far more complicated. First, the beam becomes substantially wider in the transverse direction and, second, the zone where the electrons are simultaneously reflected in the interaction region also significantly broadens. The latter circumstance indicates that the electrons are reflected back toward the injection plane from several different regions, i.e., that several VCs form in the system. An analysis of the residence times of reflected particles shows that each of the peaks in the distribution function (see Figs. 7b, 7c) corresponds to a different characteristic reflection region, i.e., to "its own" VC. In [17], it was shown that, at strong focusing magnetic fields, the chaotization of the output radiation from the vircator is due to the formation of several VCs in the drift space. Each VC affects other VCs by the electrons it reflects. As a consequence, several feedback circuits form in the system, which thus complicate the space charge dynamics of an electron beam with an overcritical current. At weak magnetic fields, the situation is analogous; however, in this case, not only electron structures (VCs) form in different cross sections along the interaction space, but also the transverse (radial) dynamics of the space charge becomes rather complicated. Hence, at weak guiding magnetic fields, the formation and interaction of electron structures in a beam with a VC substantially complicate the spectrum of the output radiation from a vircator.

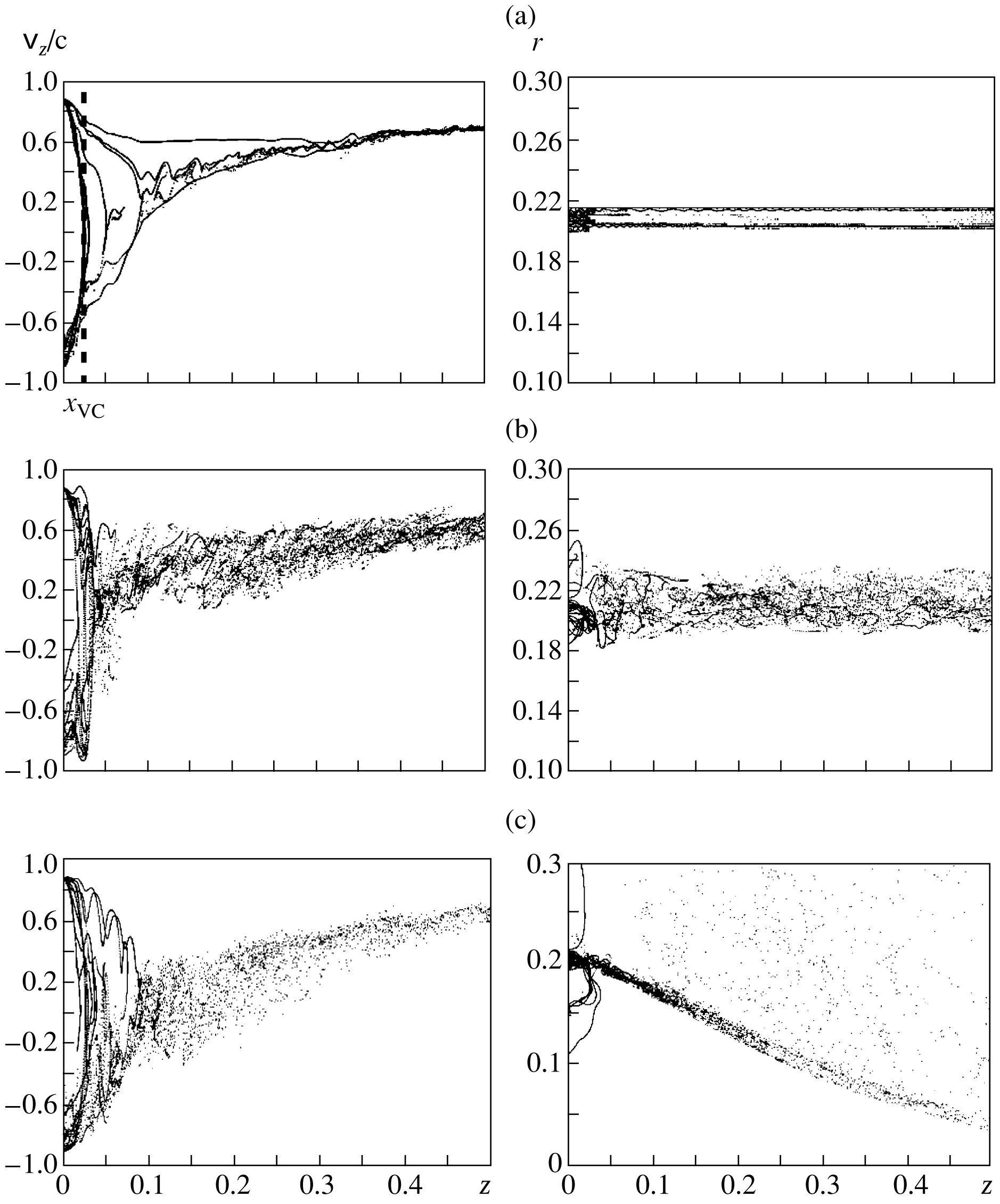

**Fig. 8.** Configurational patterns of an electron beam in the ($v_z/c$, $z$) and ($r$, $z$) coordinates in the interaction space for $B$ = (a) 40, (b) 8, and (c) 0 kG.

Note also that, as the magnetic field is decreased, the distribution function of charged particles over their residence times in the interaction space acquires a maximum at long time scales ($\tau > 6$ ns). This maximum is produced by metastable particles that remain within the VC region for a long time and that were discussed in Section 4. A comparison between Fig. 7a, 7b, and 7c, which were obtained for different magnetic field strengths $B$ leads to the conclusion that the number of metastable particles increases as the magnetic field is reduced and reaches its maximum at weak focusing magnetic fields. This conclusion agrees well with the results of numerical simulations reported in Section 4.

At the end of this section, we can draw the following conclusion: the physical mechanism whereby the dynamics of an electron beam with a VC becomes more complicated with increasing beam current or with decreasing guiding magnetic field is the formation and interaction of spatiotemporal structures in a beam with a VC. However, this mechanism manifests itself differently in different particular situations. For instance, as the guiding magnetic field is decreased, the transverse beam dynamics (two-dimensional effects) plays the main role: the transverse electron structures in a beam with an overcritical current are the first to form. As the beam current is increased at a strong focusing magnetic field, the transverse beam dynamics plays a lesser role: it manifests itself in the onset of transverse space charge oscillations.

## 6. CONCLUSIONS

In this paper, we have described the results of numerical studies of the complex oscillatory processes occurring in an REB at different strengths of the guiding magnetic field.

We have shown that, as the strength of the guiding magnetic field is decreased, the dynamics of an REB with a VC in a vircator system becomes more complicated. Our investigations demonstrate that the physical mechanism whereby the beam dynamics becomes more complicated with increasing beam current or with decreasing guiding magnetic field is the same: the formation and interaction of spatiotemporal structures in a beam with a VC. For instance, as the guiding magnetic field is decreased, the transverse beam dynamics (two-dimensional effects) plays the main role: the transverse electron structures in a beam with an overcritical current are the first to form. As the beam current is increased at a strong focusing magnetic field, the transverse beam dynamics plays a lesser role: it manifests itself merely in the onset of transverse space charge oscillations.

We have also shown that the efficiency of the interaction of the beam electrons with an electromagnetic field is very sensitive to the strength of the external magnetic field. In a vircator system, the efficiency with which an electron beam interacts with an electromagnetic field is optimum at comparatively weak focusing magnetic fields and is determined by the inherently two-dimensional motion of the beam electrons in the VC region.

## ACKNOWLEDGMENTS

We are grateful to D.I. Trubetskov for his interest in our study and valuable critical remarks and also to Yu.A. Kalinin and A.A. Koronovskiĭ for numerous fruitful discussions. This work was supported in part by the Russian Foundation for Basic Research (project nos. 05-02-16286, 05-02-08030), the Dynasty Foundation, and the International Center for Fundamental Physics in Moscow.